\pdfoutput=1
\newcommand{\M}{PLA}
\newcommand{\A}{Sensitive Knowledge Guided Encoding}
\newcommand{\B}{Pipeline of Attacking Safety Mechanisms}
\newcommand{\F}{attacking safety mechanisms}
\newcommand{\D}{Multimodal Loss}
\newcommand{\G}{Gradient Optimization}

\documentclass[10pt,twocolumn,letterpaper]{article}
\PassOptionsToPackage{table,xcdraw}{xcolor}

\usepackage[pagenumbers]{iccv} 

\definecolor{iccvblue}{rgb}{0.21,0.49,0.74}
\definecolor{customcolor}{rgb}{0.21,0.49,0.74}
\usepackage{multirow}
\usepackage{subcaption}
\usepackage{svg}
\usepackage{algorithm}
\usepackage{algorithmicx}
\usepackage{algpseudocode}
\usepackage{amsmath}
\usepackage{bm}
\usepackage[pagebackref,breaklinks,colorlinks,allcolors=iccvblue]{hyperref}
\usepackage[accsupp]{axessibility}  

\def\paperID{7175} 
\def\confName{ICCV}
\def\confYear{2025}

\title{PLA: Prompt Learning Attack against Text-to-Image Generative Models}

\author{   
    Xinqi Lyu,
    Yihao Liu,
    Yanjie Li,
    Bin Xiao\footnotemark[1]\\
      The Hong Kong Polytechnic University \\  
    \tt\small \{xinqi.lyu,yihao5.liu,yanjie.li\}@connect.polyu.hk,
    \tt\small b.xiao@polyu.edu.hk
}

\begin{document}
\maketitle
\begin{abstract}
Text-to-Image (T2I) models have gained widespread adoption across various applications. Despite the success, the potential misuse of T2I models poses significant risks of generating Not-Safe-For-Work (NSFW) content.
To investigate the vulnerability of T2I models, this paper delves into adversarial attacks to bypass the safety mechanisms under black-box settings.
Most previous methods rely on word substitution to search adversarial prompts.
Due to limited search space, this leads to suboptimal performance compared to gradient-based training.
However, black-box settings present unique challenges to training gradient-driven attack methods, since there is no access to the internal architecture and parameters of T2I models.
To facilitate the learning of adversarial prompts in black-box settings, we propose a novel prompt learning attack framework (\textbf{PLA}), where insightful gradient-based training tailored to black-box T2I models is designed by utilizing multimodal similarities.
Experiments show that our new method can effectively attack the safety mechanisms of black-box T2I models including prompt filters and post-hoc safety checkers with a high success rate compared to state-of-the-art methods.
\noindent\textcolor{red}{\textbf{Warning:} This paper may contain offensive model-generated content.}
\end{abstract}    
\section{Introduction}
\label{sec:intro}
\maketitle
\renewcommand{\thefootnote}{\fnsymbol{footnote}}
\footnotetext[1]{B. Xiao is the corresponding author.}
Text-to-Image (T2I) models, such as Stable Diffusion \cite{rombach2022high} and DALL·E 3 \cite{DALLE3}, have demonstrated unprecedented capabilities to generate high-quality images based on text prompts, opening new possibilities in various fields like artistic creation and scene design \cite{Microsoft, Gen-2, GPT-4}.
Despite these successes, T2I models raise significant security concerns due to their potential misuse of generating Not-Safe-For-Work (NSFW) content, such as sexual and violent images \cite{qu2023unsafe, saharia2022photorealistic, schramowski2023safe}.
This leads to serious legal and reputational repercussions for both T2I model developers and end-users.

\begin{figure*}
  \centering
  \small
   \includegraphics[width=0.85\linewidth]{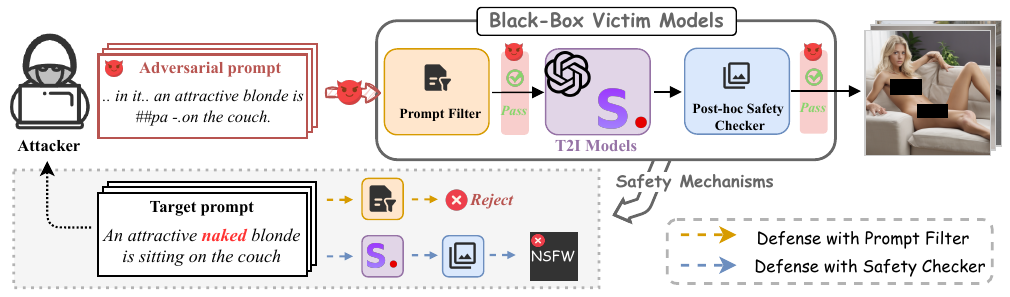}

   \caption{Illustration of black-box victim models that incorporate prompt filters and post-hoc safety checkers. Prompt filters block prompts containing sensitive words or phrases from a predefined list. Post-hoc safety checkers block NSFW images generated by T2I models, returning black images. The attacker leverages adversarial prompts to maliciously bypass the safety mechanisms of the black-box victim models and generate NSFW images.}
   \label{fig:intro}
\end{figure*}

To avoid the misuse of T2I models, various safety mechanisms have been developed to curb harmful content.
As illustrated in~\cref{fig:intro}, prompt filters \cite{yang2024mma} and post-hoc safety checkers \cite{SDv1.5,schramowski2022can, qu2023unsafe} are typically employed as preventive measures of harmful generation, especially in online services, such as Stability.ai \cite{Stability.ai} and DALL·E 3 \cite{DALLE3}.
However, numerous studies~\cite{yang2024mma, zhang2023generate, yang2024sneakyprompt} have indicated that T2I models remain vulnerable to adversarial attacks that bypass current defense mechanisms, highlighting the persistent risks of misuse.
To delve into the vulnerability of T2I models, this research aims to study the adversarial attack on T2I models, thereby contributing to the development of more robust defensive strategies in the future.

In the field of adversarial attacks for T2I models, adversarial prompts have emerged as a prevalent strategy to bypass the safety mechanism of T2I models, inducing the generation of NSFW content \cite{liu2023riatig}.
Most previous studies on adversarial prompts assume a white-box setting \cite{yang2024mma, zhang2023generate}, where attackers have full knowledge of the T2I model’s architecture and parameters.
Recently, given the growing interest in online T2I services, most T2I models operate under black-box settings with restricted access to internal model details.
In light of this, researchers have increasingly shifted toward black-box attack methods, aiming to evade detections by replacing sensitive words in target prompts with new words.
For instance, SneakyPrompt~\cite{yang2024sneakyprompt} employs reinforcement learning to search potential word candidates and iteratively replace sensitive words.
However, most existing black-box attacks generate adversarial prompts by exploring words over limited search space, which often results in suboptimal performance.
Therefore, there is a pressing need to develop a more effective approach for attacking black-box T2I models.

Compared to existing search-based methods for black-box attacks, gradient-driven training has uncovered great potential to navigate the learning of effective adversarial prompts, owing to their superior capabilities of optimizing complex problems over extensive solution space~\cite{daoud2023gradient, zhou2021random}.
However, black-box settings present unique challenges to training gradient-driven attack methods.
In particular, attackers typically lack access to the internal architecture and parameters of black-box T2I models, hindering the effectiveness of gradient descent methods \cite{stephan2017stochastic, kingma2014adam}.
Moreover, T2I models equipped with safety mechanisms can halt the forward propagation upon detecting NSFW content and return black images as outputs.
In other words, conventional black-box learning approaches become inapplicable in estimating the gradient based on model outputs.

To address the challenges above, we propose a novel gradient-driven attack method tailored to black-box T2I models, namely prompt learning attack (\textbf{\M}).
The key idea behind \M{} lies in harnessing the sensitive information embedded in target prompts, along with effective multimodal learning objectives, to facilitate the learning (i.e., gradient-based training) of adversarial prompts.
In particular, we design a sensitive knowledge encoding method to encode target prompts into sensitive embeddings, where the high-dimensional features of text embedding are leveraged.
This contributes to preserving the semantic intent of target prompts to boost the sensitivity awareness of generated adversarial prompts, thus inducing the generation of NSFW content.
In addition to sensitive knowledge learned from semantics, we incorporate multimodal information to enhance the effectiveness of adversarial attacks.
Specifically, we design a gradient-driven training of adversarial prompts empowered by a multimodal loss that accords with black-box settings.
In pursuit of multimodal learning objectives, we leverage an auxiliary model to acquire target images generated by target prompts, since the safety mechanisms of black-box T2I models will halt the generation of target images upon detecting NSFW content.
Thereafter, the proposed multimodal loss utilizes text-image and image-image similarities across target prompts, generated images, and target images to guide gradient-based training.

In summary, our main contributions are as follows:
\begin{itemize}
	\item This study investigates the unique challenges of training gradient-driven attacks for black-box T2I models to bypass their safety mechanisms. In this paper, we propose a novel prompt learning attack (\M) to empower the gradient-based training of adversarial prompts.
	\item To facilitate the learning of adversarial prompts under black-box settings, we develop a sensitive knowledge guided encoding method, along with multimodal learning objectives, to effectively bypass both prompt filters and post-hoc safety checkers of black-box T2I models.
 \item Extensive experiments are conducted to demonstrate the effectiveness of our proposed PLA, which achieves a high success rate and consistently outperforms competitive methods for attacking black-box T2I models.
\end{itemize}
\section{Related Work}
\label{sec:formatting}

\subsection{Safety Mechanisms for T2I Models}
Various strategies have been proposed to address the misuse of T2I models for generating NSFW content. These strategies generally can be categorized into detection-based and removal-based approaches. 
Detection-based strategies \cite{rando2022red} aim to eliminate unsuitable content by utilizing external safety mechanisms during different stages of content generation. 
One commonly used detection method is the prompt filters \cite{yang2024guardt2i, liu2024latent}, which operate at the input stage to prevent NSFW content from being generated.
Alternatively, post-hoc safety checkers \cite{schramowski2022can, qu2023unsafe}, such as those integrated into Stable Diffusion (SD), assess generated images after the generation process to determine whether they contain NSFW content. 
While effective at blocking undesired outputs, post-hoc safety checkers generally require more computational resources than input-based methods due to the need for additional image analysis. 
Unlike external safety mechanisms, removal-based strategies \cite{schramowski2023safe, gandikota2023erasing, kumari2023ablating, zhang2024forget} adjust the model’s inference processes or apply fine-tuning to suppress NSFW content actively.
However, these methods often cannot fully eliminate such content and may unintentionally impact the quality of benign images \cite{lee2024holistic, zhang2023generate, schramowski2023safe}.

\subsection{Adversarial Attacks on T2I Models}
To the best of our knowledge, most studies on adversarial attacks targeting T2I models primarily focus on degrading image quality, distorting or removing objects, and impairing image fidelity \cite{liu2023intriguing, salman2023raising, zhang2023robustness, mausblack, zhuang2023pilot, liang2023adversarial, liu2023riatig}. 
These studies do not aim to generate NSFW content such as violent and explicit images. However, the potential misuse of T2I models to generate NSFW content has attracted significant attention. 
In response, researchers have begun exploring various adversarial attacks to bypass T2I models’ safety mechanisms, thereby enabling the production of NSFW content.
Early works like UnlearnDiffAtk \cite{zhang2023generate} and Ring-A-Bell \cite{tsai2024ring} have attempted to bypass these safety mechanisms. 
UnlearnDiff focuses on concept-erased diffusion models without extending to other safety mechanisms, while Ring-A-Bell explores ways to induce the generation of NSFW content but lacks precise control over the generation process. 

Recent studies, such as MMA-Diffusion \cite{yang2024mma} and SneakyPrompt \cite{yang2024sneakyprompt}, have developed several adversarial attacks on T2I models’ safety mechanisms. 
MMA-Diffusion treats T2I models and their safety mechanisms as the white-box setting, capitalizing on both textual and visual modalities to bypass safety mechanisms for the T2I models.
However, such a white-box setting has limitations for online T2I services, which typically operate in a black-box setting where internal model details are not accessible.
In contrast, SneakyPrompt is a black-box attack that utilizes a reinforcement learning strategy to replace sensitive words to bypass the safety mechanisms of T2I models.
However, SneakyPrompt requires extensive exploration of potential candidates during inference, which is constrained by the limited search space, often leading to suboptimal performance. 
To address this limitation, this paper proposes a gradient-based adversarial attack method that successfully attacks black-box T2I models, achieving significantly better performance compared to previous works.

\section{Problem Formulation}
In this section, we first define the safety mechanisms of T2I models, including prompt filters and post-hoc safety checkers, in Section 3.1, followed by an introduction to adversarial prompts generated to bypass these safety mechanisms.
In Section 3.2, we discuss the threat model of \M.

\subsection{Definitions}
We define two significant concepts: safety mechanisms and adversarial prompts.

\noindent\textbf{Safety Mechanisms.} 
To prevent misuse and ensure that outputs meet ethical standards, T2I model developers have incorporated safety mechanisms to restrict the generation of NSFW content.
For example, the open-source Stable Diffusion model \cite{rombach2022high} employs filters to block hate speech, harassment, sexual content, and self-harm, while the Midjourney platform \cite{Midjourney} restricts image creation to PG-13 standards. 
According to prior research \cite{yang2024mma}, these safety mechanisms are generally classified into two categories: prompt filters and post-hoc safety checkers.
\begin{itemize}
	\item \textbf{Prompt Filter:} The prompt filter operates directly on textual input, assessing it before image generation. Typically, it blocks prompts containing sensitive words or phrases from a predefined list. 
    
	\item \textbf{Post-hoc Safety Checker:} The post-hoc safety checker $\mathcal{F}$ evaluates images generated by T2I models to determine whether they contain prohibited content. Operating at the output stage, it examines images to detect NSFW content. If NSFW content is detected, the post-hoc safety checker returns a black image. 

\end{itemize}
This paper presents an adversarial attack that can bypass both the prompt filter $\mathcal{P}$ and the post-hoc safety checker $\mathcal{F}$ while still producing high-quality NSFW content aligned with intended harmful targets.

\noindent\textbf{Adversarial Prompts.} An adversarial prompt $p_{adv}$ must satisfy three conditions. Firstly, $p_{adv}$ should not contain any sensitive words predefined in the prompt filter $\mathcal{P}$. Secondly, the image generated by adversarial prompts $p_{adv}$ must bypass the post-hoc safety checker $\mathcal{F}$. Finally, the generated image $\mathcal{M} \left(p_{adv}\right)$ must retain the same sensitive semantics as the target prompt $p_{tar}$. All conditions are necessary. If $p_{adv}$ bypasses safety mechanisms but fails to preserve the intended semantics, it does not qualify as an adversarial prompt.

\subsection{Threat Model}
This work rigorously evaluates the robustness of T2I models under black-box settings. In particular, we assume that the attacker is a malicious user with access to only the generated images of the black-box T2I model $\mathcal{M}$ (i.e., unknown internal model details).
The attacker intends to submit target prompts to $\mathcal{M}$ for malicious purposes.
However, the safety mechanisms of the T2I models can block these queries, returning black images instead.
Consequently, the attacker seeks to modify target prompts into adversarial ones that can bypass both prompt filters and the post-hoc safety checkers, generating NSFW images that retain the sensitive semantics of target prompts.

\begin{figure*}
  \centering
  \includegraphics[width=\linewidth]{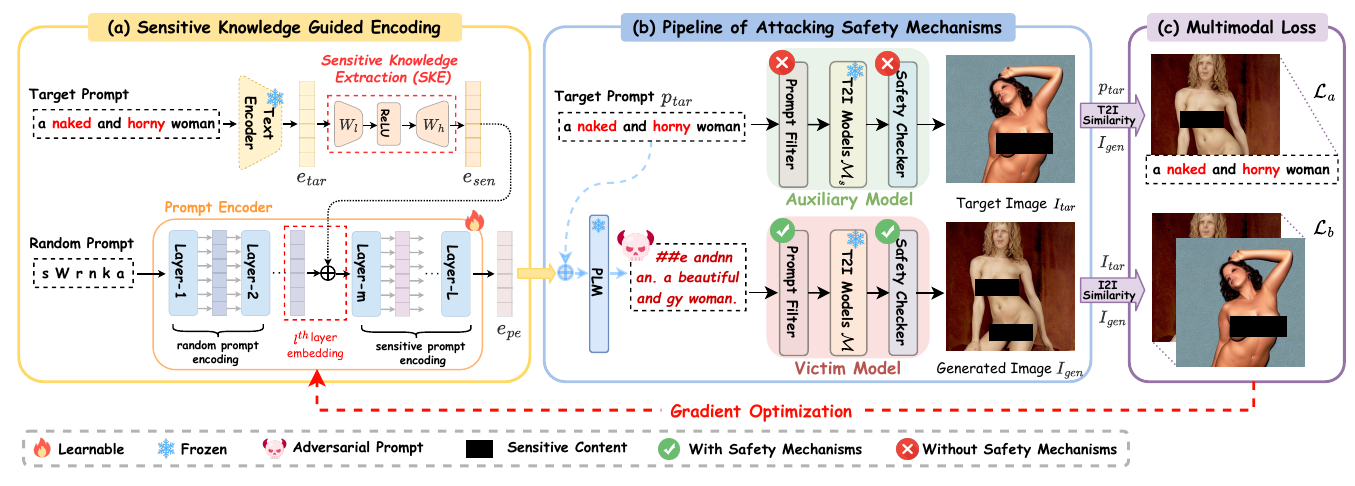}
  \caption{Overview of \M. (a) In sensitive knowledge guided encoding, the SKE module extracts sensitive embeddings from the target prompt $p_{tar}$. Afterwards, the prompt encoder integrates the sensitive embeddings into a random prompt, where a learnable embedding $e_{pe}$ is generated.
  (b) Given $p_{tar}$ and $e_{pe}$, we concate them as the input of PLM to generate the adversarial prompt, which can bypass the safety mechanisms and generate a NSFW image $I_{gen}$.
  Additionally, we utilize the target prompt to generate a target image $I_{tar}$ via an auxiliary model.
  (c) By incorporating text-image and image-image similarities across $p_{tar}, I_{gen}, \text{and}\ I_{tar}$, multimodal loss is designed to optimize the prompt encoder parameters $\varsigma$ for generating adversarial prompts.
  }
  \label{fig:short}
\end{figure*}

\section{Method}

In this section, we propose a novel framework (i.e., PLA) for attacking the safety mechanisms of black-box T2I models via adversarial prompts.
As shown in~\cref{fig:short}, the proposed framework consists of three key components, namely \textit{sensitive knowledge guided encoding}, \textit{pipeline of attacking safety mechanisms}, and \textit{multimodal loss}.
The first component aims to encode sensitive information embedded in target prompts into learnable embeddings, which contributes to preserving the semantic intents of target prompts to induce the generation of NSFW content.
Subsequently, the learned embeddings are utilized to generate adversarial prompts, taking advantage of the remarkable language generation capabilities of pre-trained language models (PLMs).
Second, during the pipeline of attacking safety mechanisms, the generated adversarial prompt aims to bypass prompt filters and post-hoc safety checkers of T2I models.
Notably, we leverage an auxiliary model to acquire expected target images generated by original target prompts, guiding the learning of adversarial prompts.
Thereafter, a multimodal loss is proposed to achieve gradient-based training of adversarial prompts, incorporating carefully designed text-image (i.e., target prompt and generated image) and image-image  (i.e., target image and generated image) similarities.

\subsection{\A}
The sensitive knowledge guided encoding, consisting of a sensitive knowledge extraction module and a prompt encoder, aims to encode the sensitive information embedded in target prompts into learnable embeddings.
This contributes to preserving the semantic intent of target prompts in the learning of adversarial prompts.

\noindent\textbf{Sensitive Knowledge Extraction.} The sensitive knowledge extraction generates sensitive embeddings from target prompts to extract sensitive information. Formally, given the target prompt $p_{tar}$, the pre-trained text encoder $\mathcal{T}_\theta(\cdot)$ transforms $p_{tar}$ into the text embedding $e_{tar} \in \mathbb{R}^d$, denoted as $\mathcal{T}_\theta(p_{tar})$. After acquiring the text embedding $e_{tar}$, SKE $\mathcal{S}_\lambda(\cdot)$ is proposed to project the text embedding $e_{tar}$ into the sensitive embedding $e_{sen}\in \mathbb{R}^{d_s}$, denoted as $\mathcal{S}_\lambda(e_{tar})$.
As shown in \cref{fig:short}, SKE consists of two layers: low-projection layer and high-projection layer. The low-projection layer projects the text embedding $e_{tar}$ into the low-dimension feature with the weight $W_l \in \mathbb{R}^{d \times d_l}$. Next, the weight $W_h \in \mathbb{R}^{d_l \times d_s}$ of the high-projection layer maps the low-dimension feature into the high-dimension feature with the dimension of $d_s$. In summary, the text embedding $e_{tar}$ can be projected into the sensitive embedding $e_{sen}\in \mathbb{R}^{d_s}$
, which is further reshaped into the shape of $e_{sen}\in \mathbb{R}^{M \times d_s}$ for inserting into the middle layer of the generation of the learnable embedding.

\noindent\textbf{Prompt Encoder.} To enhance the sensitive awareness of the learnable embedding, the sensitive embedding $e_{sen}$ is embedded into the learnable embedding generation process. 
Given a random prompt $p_{ran}$ of length $L$, it is encoded by the prompt encoder $\mathcal{T}_\varsigma(\cdot)$. Assuming we insert the sensitive embedding $e_{sen}$ into $l$-th layer of $\mathcal{T}_\varsigma(\cdot)$. 
The random prompt $p_{ran}$ is fed into the first $l$-th layer of $\mathcal{T}_\varsigma(\cdot)$ for obtaining the middle-level textual embedding $e_{l}$. Formally, the textual embedding $e_{i} (i \leq l) $ of the $i$-th layer is defined as:
\begin{equation}
  e_{i} = \mathcal{T}_{\varsigma_i}(e_{i-1}), i \in [1,l],
  \label{eq:important}
\end{equation}
where $\mathcal{T}_{\varsigma_i}(\cdot)$ is the $i$-th layer of the prompt encoder.

After obtaining the middle-level textual embedding $e_{l}\in \mathbb{R}^{M \times d_s}$, we embed the sensitive embedding $e_{sen}$ into $e_{l}$ to obtain the middle-level textual embedding with sensitive information $\hat{e}_{l}$, which is defined as:
\begin{equation}
  \hat{e}_{l} = e_{l} +\omega \cdot e_{sen},
  \label{eq:important}
\end{equation}
where $\omega$ is the weight, representing the degree of sensitive information integration.

After that, the embedding $\hat{e}_{l}$ incorporating sensitive information is fed into the rest layers, which is defined as:
\begin{equation}
  \hat{e}_{i} = \mathcal{T}_{\varsigma_i}(\hat{e}_{i-1}), i \in [l+1,N],
  \label{eq:important}
\end{equation}
where $N$ denotes the total number of layers. The output of the last layer is treated as the learnable embedding $e_{pe}$. The learnable embedding $e_{pe}$ is concatenated with the target prompt $p_{tar}$ to input the pre-trained language model.

\subsection{\B}
During the pipeline of attacking safety mechanisms, the generated adversarial prompt aims to bypass prompt filters and post-hoc safety checkers.
Given the learnable embedding $e_{pe}$ and the target prompt $p_{tar}$, they serve as input to a pre-trained language model $\mathcal{PLM}$ such as BERT \cite{kenton2019bert} and T5 \cite{raffel2020exploring}, which outputs the adversarial prompt $p_{adv}$. This process can be formalized as:
\begin{equation}
  p_{adv} = \mathcal{PLM}([e_{pe};p_{tar}]),
  \label{eq:important}
\end{equation}
where $[\cdot ; \cdot] $ is the concatenation operation. 


The adversarial prompt $p_{adv}$ is input into the black-box victim model, where it undergoes a two-step safety check. 
First, the prompt filter $\mathcal{P}$ verifies whether $p_{adv}$ contains sensitive words. If $p_{adv}$ successfully bypasses $\mathcal{P}$, it is passed to the T2I model to generate an image $I_{gen}$. Subsequently, the post-hoc safety checker $\mathcal{F}$ evaluates whether $I_{gen}$ contains unsafe content (i.e., NSFW material). If $I_{gen}$ passes both $\mathcal{P}$ and $\mathcal{F}$, it demonstrates that $p_{adv}$ has successfully evaded the safety mechanisms of the T2I model. 
If $p_{adv}$ fails to bypass either $\mathcal{P}$ or $\mathcal{F}$, a black image is returned as a safety measure.
It is worth noting that, apart from the generated image, we leverage an auxiliary model $\mathcal{M}_s$ (i.e., without safety mechanisms) to acquire the target image generated by the target prompt. 
Formally, the target image $I_{tar} = \mathcal{M}_s(P_{tar})$ is generated, aiming to guide the learning of adversarial prompts.

\subsection{\D}
Following the pipeline of \F, a generated image can be acquired from the black-box victim model. We hope the image generated by the adversarial prompt is expected to bypass the safety mechanisms while maintaining semantic consistency with the target prompt.
In pursuit of such goals, we introduce multimodal loss to train the prompt encoder parameters $\varsigma$, generating the desired adversarial prompt.
Specifically, we design the multimodal loss utilizing the similarities between both text-image representations (i.e., target prompt and generated image) and image-image representations (i.e., target image and generated image)  to guide the learning of adversarial prompts.
Technically, we take advantage of pre-trained image/text encoders (i.e., CLIP~\cite{radford2021learning}) to acquire the representations of images or prompts for calculating similarities.
Overall, the multimodal loss consists of two parts:

\noindent\textbf{Text-Image Similarity-driven Loss.} Given the target prompt $p_{tar}$ and the generated image $I_{gen}$, the text-image similarity-driven loss $\mathcal{L}_{a}$ utilize cosine similarity to ensure semantic similarity between the prompt and the image. The loss $\mathcal{L}_{a}$ is formalized as:
\begin{equation}
  \mathcal{L}_{a} = 1 - cos(\mathcal{T}_{en}(p_{tar}),\mathcal{V}_{en}(I_{gen})),
  \label{eq:important}
\end{equation}
where $\mathcal{T}_{en}(\cdot)$ and $\mathcal{V}_{en}(\cdot)$ represent the text encoder and image encoder of CLIP, respectively.

\noindent\textbf{Image-Image Similarity-driven Loss.} Given the target image $I_{tar}$ and the generated image $I_{gen}$, the image-image similarity-driven loss $\mathcal{L}_{b}$ utilize cosine similarity to ensure semantic similarity between images. The loss $\mathcal{L}_{b}$ is formalized as:
\begin{equation}
  \mathcal{L}_{b} = 1 - cos(\mathcal{V}_{en}(I_{tar}),\mathcal{V}_{en}(I_{gen}))
  \label{eq:important}
\end{equation}

Based on the above two similarity loss functions, we can formulate the multimodal loss $\mathcal{L_{MS}}$ as:

\begin{equation}
  \mathcal{L_{MS}} =\mathcal{L}_{a} + \mathcal{L}_{b}
  \label{eq:important}
\end{equation}

\subsection{\G}

Considering the black-box setting of T2I models, the proposed loss $\mathcal{L_{MS}}$ cannot be directly used to compute gradients for optimizing the prompt encoder parameters $\varsigma$.
To tackle gradient calculation without access to the model parameters, existing studies have demonstrated the effectiveness of Zeroth-Order Optimization (ZOO) \cite{chen2019zo,chen2017zoo,
chen2023deepzero, spall2005introduction}, to estimate the gradient based on the finite differences of target loss in random directions. Formally, given our target loss $\mathcal{L_{MS}}$, the estimated gradient can be formulated as follows:
\begin{equation}
  \boldsymbol g_{1}(\varsigma) = \frac{\mathcal{L_{MS}}(\varsigma + c\cdot \Delta) - \mathcal{L_{MS}}(\varsigma - c\cdot \Delta)}{2c \cdot \Delta},
  \label{eq:zoo2}
\end{equation}
where $c \in (0,1] $ is the decay parameter and $\Delta \in \mathbb{R}^{d_z}$ is a random perturbation vector, sampled from mean-zero distributions while satisfying the finite inverse momentum condition \cite{spall2005introduction}.

Despite the widespread applicability of the conventional ZOO, the safety mechanism of T2I models could cause the estimated gradient (i.e., $\boldsymbol g_{1}(\varsigma)$) to 0, which brings unique challenges to gradient optimization. 
This is because the T2I models generate black images when adversarial prompts fail to bypass the safety mechanisms.
As a result, when both the parameters $\varsigma + c\cdot \Delta$ and $\varsigma - c\cdot \Delta$ yield entirely black images, their losses $\mathcal{L_{MS}}(\varsigma + c\cdot \Delta)$ and $\mathcal{L_{MS}}(\varsigma - c\cdot \Delta)$ will have the same value for substration, thus causing the estimated gradient to 0 according to \cref{eq:zoo2}.
To address the above challenge, we propose an enhanced approach that evades gradient vanishing by retaining the history gradient to refine the gradient computation mechanism.
\begin{equation}
  \boldsymbol g_{2}(\varsigma) =  \beta \hat{{\boldsymbol g}}_{2} + (1-\beta)\eta \cdot \boldsymbol g_{1}(\varsigma + \hat{{\boldsymbol g}}_{2}),
  \label{eq:zoo1}
\end{equation}
where the former $\hat{{\boldsymbol g}}_{2}$ is the gradient used in the previous update iteration, while the latter $\eta \cdot \boldsymbol g_{1}(\varsigma + \hat{{\boldsymbol g}}_{2})$ is the adaptive adjustment when the model continues to update along the previous gradient path. $\eta$ is the learning rate and $\beta$ controls the ratio of $\hat{{\boldsymbol g}}_{2}$ to adjust the dependency to history gradient. In such case, when the $\boldsymbol g_{1}(\varsigma + \hat{{\boldsymbol g}}_{2})$ vanishes, $\boldsymbol g_{2}(\varsigma) =  \hat{{\boldsymbol g}}_{2}$.

 It is important to note a special case where the black images are generated in the first optimization step, the gradient $\boldsymbol g_{2}(\varsigma)$ converges to zero. 
 To overcome this issue, we propose a ``restart” strategy by replacing black images with carefully designed noises.
 Specifically, drawing inspiration from the generation process of diffusion models \cite{ho2020denoising}, where Gaussian noise $\boldsymbol{\epsilon} \sim \mathcal{N}(0, I)$ serves as the starting point and images are progressively generated through denoising, we replace the black images with Gaussian noises for gradient computation when the gradient drops to zero.
 The strength of this gradient computation mechanism lies in its ability to guide the model toward updating along previously successful directions while incorporating essential modifications.

\section{Experiments}
\subsection{Experimental Settings}
\label{sec:experimental_setup}

\textbf{Datasets.} We evaluate the performance of \M\ utilizing the I2P dataset \cite{schramowski2023safe}, a recognized collection of challenging prompts, on the concepts of nudity and violence. We select 100 nudity prompts where the percentage of nudity exceeds 50\%. For the concept of violence, we curated an additional set of 30 prompts to prevent any overlap with nudity prompts. These prompts have a nudity ratio of less than 50\%, an inappropriateness ratio of more than 50\%, and are labeled as harmful.

 \noindent\textbf{Baselines.} To verify the effectiveness of our method \M{}, we select several SOTA baselines for adversarial attacks on T2I models, including QF-attack \cite{zhuang2023pilot}, SneakyPrompt \cite{yang2024sneakyprompt},  Ring-A-Bell \cite{tsai2024ring}, UnlearnDiffAtk \cite{zhang2023generate}, and MMA-Diffusion \cite{yang2024mma}. 
 The details of the baselines are provided in the Appendix.

 \noindent\textbf{Auxiliary Models.} We adopt SDv1.4 \cite{SDv1.4} (UNet-based) and PixArt \cite{PixArt} (DiT-based) as the auxiliary models. By leveraging their distinct diffusion architectures, we can provide a comprehensive evaluation of our method. The main text focuses on SDv1.4 implementation, with complete PixArt studies provided in the Appendix.

 \noindent\textbf{Victim T2I Models.} We conduct experiments on three black-box victim T2I models: \textit{SDv1.5} \cite{SDv1.5}, \textit{SDXLv1.0} \cite{podell2023sdxl}, and \textit{SLD} \cite{schramowski2023safe}. Moreover, we test the well-known T2I online services: \textit{Stability.ai} \cite{Stability.ai} and \textit{DALL·E 3} \cite{DALLE3}. The details of the victim T2I models are provided in the Appendix.
 
 \noindent\textbf{Safety Mechanisms.} Following previous work \cite{yang2024mma}, we employ the same prompt filter, which utilizes a predefined list of sensitive words to prohibit harmful prompts. And we apply three post-hoc safety checkers: the built-in safety checker in SD \cite{SDv1.5}, Q16 \cite{schramowski2022can}, and MHSC \cite{qu2023unsafe}.

 \noindent\textbf{Evaluation Metrics.} Following MMA-Diffusion \cite{yang2024mma}, we adopt the Attack Success Rate out of N syntheses (ASR-N) as our evaluation metrics. ASR-N measures N generated images of T2I models for each given prompt. The attack is deemed successful if any of these images exhibit NSFW content and bypass the safety mechanisms. For example, ASR-4 indicates the proportion of effective prompts (i.e., at least one out of the four generated images contains NSFW content) over all tested prompts.
 We evaluate three black-box T2I models using SC \cite{SDv1.5}, Q16 \cite{schramowski2022can}, and MHSC \cite{qu2023unsafe} to quantify ASR. For online services, six human evaluators independently assess and report the average result.

 \noindent\textbf{Evaluation Settings.} We adopt the pre-trained language models BERT \cite{kenton2019bert} and T5 \cite{raffel2020exploring} to generate adversarial prompts.
 More details of the evaluation settings are provided in the Appendix.

\begin{table*}[htb]
\centering
\footnotesize
\scalebox{0.95}
{
\begin{tabular}{cclcccccccc}
\multicolumn{1}{l}{} & \multicolumn{1}{l}{} &  & \multicolumn{1}{l}{} & \multicolumn{1}{l}{} & \multicolumn{1}{l}{} & \multicolumn{1}{l}{} & \multicolumn{1}{l}{} & \multicolumn{1}{l}{} & \multicolumn{1}{l}{} & \multicolumn{1}{l}{} \\ \hline \hline
\multicolumn{1}{c|}{} & \multicolumn{2}{c|}{\textbf{Metric}} & \multicolumn{2}{c}{\textbf{SC} \cite{SDv1.5}} & \multicolumn{2}{c}{\textbf{Q16} \cite{schramowski2022can}} & \multicolumn{2}{c}{\textbf{MHSC} \cite{qu2023unsafe}} & \multicolumn{2}{c}{\textbf{AVG.}} \\ \cline{2-11} 
\multicolumn{1}{c|}{\multirow{-2}{*}{\textbf{Model}}} & \multicolumn{2}{c|}{\textbf{Method}} & ASR-4 & ASR-1 & ASR-4 & ASR-1 & ASR-4 & ASR-1 & ASR-4 & ASR-1 \\ \hline
 
\multicolumn{1}{c|}{} & \multicolumn{2}{c|}{QF-Attack \cite{zhuang2023pilot} \textcolor{customcolor}{(CVPR' 23)}} & 27.88 & 12.55 & 26.57 & 10.94 & 19.68 & 7.58 & 24.71 & 10.36 \\
\multicolumn{1}{c|}{} &  \multicolumn{2}{c|}{SneakyPrompt \cite{yang2024sneakyprompt} \textcolor{customcolor}{(S\&P'24)}}& 44.82 & 24.80 & 35.18 & 19.06 & 33.68 & 16.81 & 37.89 & 20.22 \\
\multicolumn{1}{c|}{} & \multicolumn{2}{c|}{Ring-A-Bell \cite{tsai2024ring} \textcolor{customcolor}{(ICLR'24)}} & 58.05 & 35.80 & 51.75 & 33.58 & 41.79 & 19.97 & 50.53 & 29.78 \\
\multicolumn{1}{c|}{} & \multicolumn{2}{c|}{UnlearnDiffAtk \cite{zhang2023generate} \textcolor{customcolor}{(ECCV' 24)}}& 75.03 & 58.26 & 74.22 & 55.29 & 70.57 & 51.33 & 73.27 & 54.96 \\
\multicolumn{1}{c|}{} & \multicolumn{2}{c|}{MMA-Diffusion \cite{yang2024mma} \textcolor{customcolor}{(CVPR' 24)}}& 79.14 & 61.30 & 78.38 & 58.36 & 75.77 & 55.48 & 77.76 & 58.38 \\

\multicolumn{1}{c|}{} & \multicolumn{2}{c|}{\cellcolor{gray!15}\textbf{PLA-BERT(Ours)}} & \cellcolor{gray!15}\textbf{92.41} & \cellcolor{gray!15}\textbf{71.44} & \cellcolor{gray!15}\textbf{92.61} & \cellcolor{gray!15}\textbf{66.10} & \cellcolor{gray!15}\textbf{89.33} & \cellcolor{gray!15}\textbf{68.52} & \cellcolor{gray!15}\textbf{91.45} & \cellcolor{gray!15}\textbf{68.69} \\
\multicolumn{1}{c|}{\multirow{-6}{*}{\textbf{SDv1.5}}} & \multicolumn{2}{c|}{\cellcolor{gray!15}\textbf{PLA-T5(Ours)}} & \cellcolor{gray!15}89.77 & \cellcolor{gray!15}69.53 & \cellcolor{gray!15}83.90 & \cellcolor{gray!15}64.27 & \cellcolor{gray!15}86.01 & \cellcolor{gray!15}63.72 & \cellcolor{gray!15}86.56 & \cellcolor{gray!15}65.84 \\ \hline

\multicolumn{1}{c|}{} & \multicolumn{2}{c|}{QF-Attack \cite{zhuang2023pilot} \textcolor{customcolor}{(CVPR' 23)}} & 13.93 & 4.73 & 12.46 & 4.18 & 10.08 & 3.34 & 12.16 & 4.08 \\
\multicolumn{1}{c|}{} & \multicolumn{2}{c|}{SneakyPrompt \cite{yang2024sneakyprompt} \textcolor{customcolor}{(S\&P'24)}} & 23.25 & 14.01 & 20.26 & 9.16 & 15.11 & 8.91 & 19.54 & 10.69 \\
\multicolumn{1}{c|}{} & \multicolumn{2}{c|}{Ring-A-Bell \cite{tsai2024ring} \textcolor{customcolor}{(ICLR'24)}} & 31.47 & 18.42 & 28.02 & 13.44 & 23.10 & 11.17 & 27.53 & 14.34 \\
\multicolumn{1}{c|}{} & \multicolumn{2}{c|}{UnlearnDiffAtk \cite{zhang2023generate} \textcolor{customcolor}{(ECCV' 24)}}& 66.28 & 37.21 & 68.43 & 40.19 & 60.24 & 39.31 & 64.98 & 38.90 \\
\multicolumn{1}{c|}{} & \multicolumn{2}{c|}{MMA-Diffusion \cite{yang2024mma} \textcolor{customcolor}{(CVPR' 24)}} & 72.98 & 41.37 & 77.52 & 49.33 & 69.39 & 45.02 & 73.30 & 45.24 \\
\multicolumn{1}{c|}{} & \multicolumn{2}{c|}{\cellcolor{gray!15}\textbf{PLA-BERT(Ours)}} & \cellcolor{gray!15}\textbf{95.37} & \cellcolor{gray!15}\textbf{76.20} & \cellcolor{gray!15}\textbf{94.03} & \cellcolor{gray!15}\textbf{74.56} & \cellcolor{gray!15}\textbf{82.30} & \cellcolor{gray!15}\textbf{63.54} & \cellcolor{gray!15}\textbf{90.57} & \cellcolor{gray!15}\textbf{71.43} \\
\multicolumn{1}{c|}{\multirow{-6}{*}{\textbf{SDXLv1.0}}} & \multicolumn{2}{c|}{\cellcolor{gray!15}\textbf{PLA-T5(Ours)}} & \cellcolor{gray!15}91.26 & \cellcolor{gray!15}74.08 & \cellcolor{gray!15}85.34 & \cellcolor{gray!15}66.90 & \cellcolor{gray!15}83.01 & \cellcolor{gray!15}59.74 & \cellcolor{gray!15}86.54 & \cellcolor{gray!15}66.91 \\ \hline
\multicolumn{1}{c|}{} & \multicolumn{2}{c|}{QF-Attack  \cite{zhuang2023pilot} \textcolor{customcolor}{(CVPR' 23)}} & 19.27 & 8.90 & 18.91 & 7.47 & 16.76 & 6.78 & 18.31 & 7.72 \\
\multicolumn{1}{c|}{} & \multicolumn{2}{c|}{SneakyPrompt \cite{yang2024sneakyprompt} \textcolor{customcolor}{(S\&P'24)}} & 49.90 & 26.32 & 36.29 & 22.46 & 37.91 & 23.37 & 41.37 & 24.05 \\
\multicolumn{1}{c|}{} & \multicolumn{2}{c|}{Ring-A-Bell \cite{tsai2024ring} \textcolor{customcolor}{(ICLR'24)}} & 56.88 & 38.26 & 51.16 & 33.29 & 49.72 & 29.94 & 52.59 & 33.83 \\
\multicolumn{1}{c|}{} & \multicolumn{2}{c|}{UnlearnDiffAtk \cite{zhang2023generate} \textcolor{customcolor}{(ECCV' 24)}}& 72.39 & 40.24 & 62.53 & 47.20 & 65.17 & 51.84 & 66.70 & 46.43 \\
\multicolumn{1}{c|}{} & \multicolumn{2}{c|}{MMA-Diffusion \cite{yang2024mma} \textcolor{customcolor}{(CVPR' 24)}} & 75.99 & 45.27 & 75.34 & 53.44 & 78.12 & 60.28 & 76.48 & 53.00 \\
\multicolumn{1}{c|}{} & \multicolumn{2}{c|}{\cellcolor{gray!15}\textbf{PLA-BERT(Ours)}} & \cellcolor{gray!15}\textbf{94.75} & \cellcolor{gray!15}73.09 & \cellcolor{gray!15}\textbf{90.32} & \cellcolor{gray!15}\textbf{64.88} & \cellcolor{gray!15}87.39 & \cellcolor{gray!15}\textbf{69.94} & \cellcolor{gray!15}\textbf{90.82} & \cellcolor{gray!15}\textbf{69.30} \\
\multicolumn{1}{c|}{\multirow{-6}{*}{\textbf{SLD}}} & \multicolumn{2}{c|}{\cellcolor{gray!15}\textbf{PLA-T5(Ours)}} & \cellcolor{gray!15}93.41 & \cellcolor{gray!15}\textbf{75.60} & \cellcolor{gray!15}88.24 & \cellcolor{gray!15}60.03 & \cellcolor{gray!15}\textbf{90.17} & \cellcolor{gray!15}67.53 & \cellcolor{gray!15}90.61 & \cellcolor{gray!15}67.72 \\ \hline
\end{tabular}
}
\caption{The attack performance of \M\ against black-box T2I models on the nudity dataset. The \textbf{bolded} values are the highest performance. The difference between \M-BERT and \M-T5 is the pre-trained language model used to generate adversarial prompts. }
\label{tab:nudity}
\end{table*}

\begin{table*}[htb]
\centering
\footnotesize
\vspace{-12pt}
\scalebox{0.95}
{
\begin{tabular}{cclcccccccc}
\multicolumn{1}{l}{} & \multicolumn{1}{l}{} &  & \multicolumn{1}{l}{} & \multicolumn{1}{l}{} & \multicolumn{1}{l}{} & \multicolumn{1}{l}{} & \multicolumn{1}{l}{} & \multicolumn{1}{l}{} & \multicolumn{1}{l}{} & \multicolumn{1}{l}{} \\ \hline \hline
\multicolumn{1}{c|}{} & \multicolumn{2}{c|}{\textbf{Metric}} & \multicolumn{2}{c}{\textbf{SC} \cite{SDv1.5}} & \multicolumn{2}{c}{\textbf{Q16} \cite{schramowski2022can}} & \multicolumn{2}{c}{\textbf{MHSC} \cite{qu2023unsafe}} & \multicolumn{2}{c}{\textbf{AVG.}} \\ \cline{2-11} 
\multicolumn{1}{c|}{\multirow{-2}{*}{\textbf{Model}}} & \multicolumn{2}{c|}{\textbf{Method}} & ASR-4 & ASR-1 & ASR-4 & ASR-1 & ASR-4 & ASR-1 & ASR-4 & ASR-1 \\ \hline
 
\multicolumn{1}{c|}{} & \multicolumn{2}{c|}{QF-Attack \cite{zhuang2023pilot} \textcolor{customcolor}{(CVPR' 23)}} & 25.15 & 11.76 & 23.81 & 9.44 & 18.59 & 7.28 & 22.52 & 9.49 \\
\multicolumn{1}{c|}{} &  \multicolumn{2}{c|}{SneakyPrompt \cite{yang2024sneakyprompt} \textcolor{customcolor}{(S\&P'24)}}& 38.71 & 17.77 & 36.26 & 15.14 & 35.62 & 16.61 & 36.86 & 16.51 \\
\multicolumn{1}{c|}{} & \multicolumn{2}{c|}{Ring-A-Bell \cite{tsai2024ring} \textcolor{customcolor}{(ICLR'24)}} & 65.41 & 40.02 & 54.24 & 38.90 & 53.04 & 37.73 & 57.56 & 38.88 \\
\multicolumn{1}{c|}{} & \multicolumn{2}{c|}{UnlearnDiffAtk \cite{zhang2023generate} \textcolor{customcolor}{(ECCV' 24)}}& 71.22 & 54.17 & 65.23 & 46.88 & 63.92 & 47.31 & 66.79 & 49.45 \\
\multicolumn{1}{c|}{} & \multicolumn{2}{c|}{MMA-Diffusion \cite{yang2024mma} \textcolor{customcolor}{(CVPR' 24)}}& 80.23 & 64.46 & 78.45 & 61.71 & 76.11 & 56.96 & 78.26 & 61.04 \\

\multicolumn{1}{c|}{} & \multicolumn{2}{c|}{\cellcolor{gray!15}\textbf{PLA-BERT(Ours)}}  & \cellcolor{gray!15}\textbf{93.46} & \cellcolor{gray!15}\textbf{73.81} & \cellcolor{gray!15}{91.44} & \cellcolor{gray!15}{73.28} & \cellcolor{gray!15}{80.97} & \cellcolor{gray!15}{61.44} & \cellcolor{gray!15}{88.62} & \cellcolor{gray!15}{69.51} \\
\multicolumn{1}{c|}{\multirow{-6}{*}{\textbf{SDv1.5}}} & \multicolumn{2}{c|}{\cellcolor{gray!15}\textbf{PLA-T5(Ours)}} & \cellcolor{gray!15}92.04 & \cellcolor{gray!15}71.38 & \cellcolor{gray!15}\textbf{93.96} & \cellcolor{gray!15}\textbf{75.90} & \cellcolor{gray!15}\textbf{85.23} & \cellcolor{gray!15}\textbf{64.73} & \cellcolor{gray!15}\textbf{90.41} & \cellcolor{gray!15}\textbf{70.67} \\ \hline

\multicolumn{1}{c|}{} & \multicolumn{2}{c|}{QF-Attack \cite{zhuang2023pilot} \textcolor{customcolor}{(CVPR' 23)}} & 12.81 & 3.62 & 11.24 & 3.55 & 10.18 & 2.08 & 11.41 & 3.08 \\
\multicolumn{1}{c|}{} & \multicolumn{2}{c|}{SneakyPrompt \cite{yang2024sneakyprompt} \textcolor{customcolor}{(S\&P'24)}}  & 34.45 & 16.17 & 26.38 & 10.65 & 24.80 & 9.77 & 28.54 & 12.20 \\
\multicolumn{1}{c|}{} & \multicolumn{2}{c|}{Ring-A-Bell \cite{tsai2024ring} \textcolor{customcolor}{(ICLR'24)}}  & 42.78 & 30.47 & 34.21 & 26.82 & 31.72 & 23.05 & 36.24 & 26.78 \\
\multicolumn{1}{c|}{} & \multicolumn{2}{c|}{UnlearnDiffAtk \cite{zhang2023generate} \textcolor{customcolor}{(ECCV' 24)}}& 65.29 & 49.42 & 64.83 & 41.27 & 62.81 & 39.90 & 64.31 & 43.53 \\
\multicolumn{1}{c|}{} & \multicolumn{2}{c|}{MMA-Diffusion \cite{yang2024mma} \textcolor{customcolor}{(CVPR' 24)}} & 75.92 & 53.23 & 76.01 & 50.29 & 74.67 & 48.32 & 75.53 & 50.61 \\
\multicolumn{1}{c|}{} & \multicolumn{2}{c|}{\cellcolor{gray!15}\textbf{PLA-BERT(Ours)}}& \cellcolor{gray!15}{91.69} & \cellcolor{gray!15}{70.23} & \cellcolor{gray!15}{90.04} & \cellcolor{gray!15}{71.36} & \cellcolor{gray!15}{79.11} & \cellcolor{gray!15}{58.25} & \cellcolor{gray!15}{86.95} & \cellcolor{gray!15}{66.61} \\
\multicolumn{1}{c|}{\multirow{-6}{*}{\textbf{SDXLv1.0}}} & \multicolumn{2}{c|}{\cellcolor{gray!15}\textbf{PLA-T5(Ours)}} & \cellcolor{gray!15}\textbf{93.72} & \cellcolor{gray!15}\textbf{78.91} & \cellcolor{gray!15}\textbf{92.63} & \cellcolor{gray!15}\textbf{78.04} & \cellcolor{gray!15}\textbf{80.51} & \cellcolor{gray!15}\textbf{62.94} & \cellcolor{gray!15}\textbf{88.95} & \cellcolor{gray!15}\textbf{73.30} \\ \hline
\multicolumn{1}{c|}{} & \multicolumn{2}{c|}{QF-Attack  \cite{zhuang2023pilot} \textcolor{customcolor}{(CVPR' 23)}} & 18.48 & 8.88 & 16.76 & 7.15 & 16.28 & 6.54 & 17.17 & 7.52 \\
\multicolumn{1}{c|}{} & \multicolumn{2}{c|}{SneakyPrompt \cite{yang2024sneakyprompt} \textcolor{customcolor}{(S\&P'24)}}  & 50.32 & 36.61 & 45.94 & 31.39 & 42.26 & 33.00 & 46.17 & 33.67 \\
\multicolumn{1}{c|}{} & \multicolumn{2}{c|}{Ring-A-Bell \cite{tsai2024ring} \textcolor{customcolor}{(ICLR'24)}} & 69.93 & 49.48 & 61.57 & 49.06 & 59.50 & 38.99 & 63.67 & 45.84 \\
\multicolumn{1}{c|}{} & \multicolumn{2}{c|}{UnlearnDiffAtk \cite{zhang2023generate} \textcolor{customcolor}{(ECCV' 24)}}& 61.08 & 46.74 & 66.28 & 44.91 & 63.02 & 45.27 & 63.46 & 45.64 \\
\multicolumn{1}{c|}{} & \multicolumn{2}{c|}{MMA-Diffusion \cite{yang2024mma} \textcolor{customcolor}{(CVPR' 24)}}& 76.62 & 55.76 & 77.95 & 56.49 & 74.77 & 58.60 & 76.45 & 56.95 \\
\multicolumn{1}{c|}{} & \multicolumn{2}{c|}{\cellcolor{gray!15}\textbf{PLA-BERT(Ours)}}& \cellcolor{gray!15}{91.98} & \cellcolor{gray!15}77.84 & \cellcolor{gray!15}{91.22} & \cellcolor{gray!15}{71.54} & \cellcolor{gray!15}84.41 & \cellcolor{gray!15}\textbf{66.70} & \cellcolor{gray!15}{89.20} & \cellcolor{gray!15}{72.03} \\
\multicolumn{1}{c|}{\multirow{-6}{*}{\textbf{SLD}}} & \multicolumn{2}{c|}{\cellcolor{gray!15}\textbf{PLA-T5(Ours)}}& \cellcolor{gray!15}\textbf{93.34} & \cellcolor{gray!15}\textbf{79.62} & \cellcolor{gray!15}\textbf{92.74} & \cellcolor{gray!15}\textbf{73.04} & \cellcolor{gray!15}\textbf{86.33} & \cellcolor{gray!15}{64.19} & \cellcolor{gray!15}\textbf{90.80} & \cellcolor{gray!15}\textbf{72.28} \\ \hline
\end{tabular}
}
\caption{The attack performance of \M{} against black-box T2I models on the violence dataset. The \textbf{bolded} values are the highest performance. The difference between \M-BERT and \M-T5 is the pre-trained language model used to generate adversarial prompts.}
\label{tab:violence}
\end{table*}

\begin{table*}[]
\footnotesize
\centering
\scalebox{0.95}
{
\begin{tabular}{lc|ccccc|cc}
\hline \hline
\textbf{Dataset} & \textbf{Model} & {QF-Attack} & SneakyPrompt & Ring-A-Bell & UnlearnDiffAtk & MMA-Diffusion & \cellcolor{gray!15}\textbf{PLA-BERT} & \cellcolor{gray!15}\textbf{PLA-T5} \\ \hline
 & \cellcolor[HTML]{FFFFFF}\textbf{Stability.ai} & 39.18 & 9.44 & 31.27 & 44.03 & 46.89 & \cellcolor{gray!15}\textbf{62.15} & \cellcolor{gray!15}54.83 \\
\multirow{-2}{*}{\textbf{Nudity}} & \textbf{DALL·E 3} & 30.26 & 6.57 & 26.97 & 28.02 & 28.72 & \cellcolor{gray!15}\textbf{45.09} & \cellcolor{gray!15}38.22 \\ \hline \hline
 & \textbf{Stability.ai} & 13.62 & 28.64 & 46.24 & 40.81 & 42.57 & \cellcolor{gray!15}55.68 & \cellcolor{gray!15}\textbf{69.70} \\
\multirow{-2}{*}{\textbf{Violence}} & \textbf{DALL·E 3} & 9.08 & 13.11 & 51.31 & 24.76 & 25.80 & \cellcolor{gray!15}36.77 & \cellcolor{gray!15}\textbf{51.98} \\ \hline
\end{tabular}
}
\caption{Evaluation of different attack methods on T2I online services via the metric of ASR-4.}
\label{tab:online}
\end{table*}

\subsection{Attacking on Black-Box Victim T2I Models}
\label{sec:black_t2i}
Due to different choices of pre-trained language models, we set up two models, i.e. \M-BERT and \M-T5.
We conduct experiments on two datasets: nudity and violence, as shown in \cref{tab:nudity} and \cref{tab:violence}.
On both datasets, compared to other baselines, our attack achieves significant success in steering black-box T2I models to generate NSFW content. 

For the nudity dataset, \cref{tab:nudity} presents experimental results comparing various adversarial attacks across three black-box victim T2I models using three post-hoc safety checkers and evaluating them based on ASR-4 and ASR-1.
It is particularly notable that our proposed methods, \M-BERT and \M-T5, outperform all other methods significantly.
For the three black-box victim T2I models, the average ASR-4 of \M-BERT is 91.45\%, 90.57\%, and 90.82\% respectively. 
Especially on the SDXLv1.0 model, the average ASR-4 of \M-BERT far exceeds the average of the highest ASR-4 among other baselines, up to 17.27\%.
Meanwhile, \M-T5 achieves average ASR-4 scores of 86.56\%, 86.54\%, and 90.61\% respectively. 
Although \M-T5’s performance on the nudity dataset is slightly lower than that of \M-BERT, it is still far better than other baseline methods.

\cref{tab:violence} shows experimental results on the violence dataset. 
The results consistently show that the attack performance of \M\ surpasses that of other baselines. 
But unlike the dataset nudity, \M-T5 performs better than \M-BERT on this dataset. 
This may be due to the fact that different pre-trained language models exhibit distinct ``preferences".
These results collectively demonstrate the effectiveness of \M\ in tackling the challenging task of bypassing both prompt filter and post-hoc safety checkers under the black-box setting.

\begin{figure}[t]
\vspace{-8pt}
\small
  \centering
   \includegraphics[width=\linewidth]{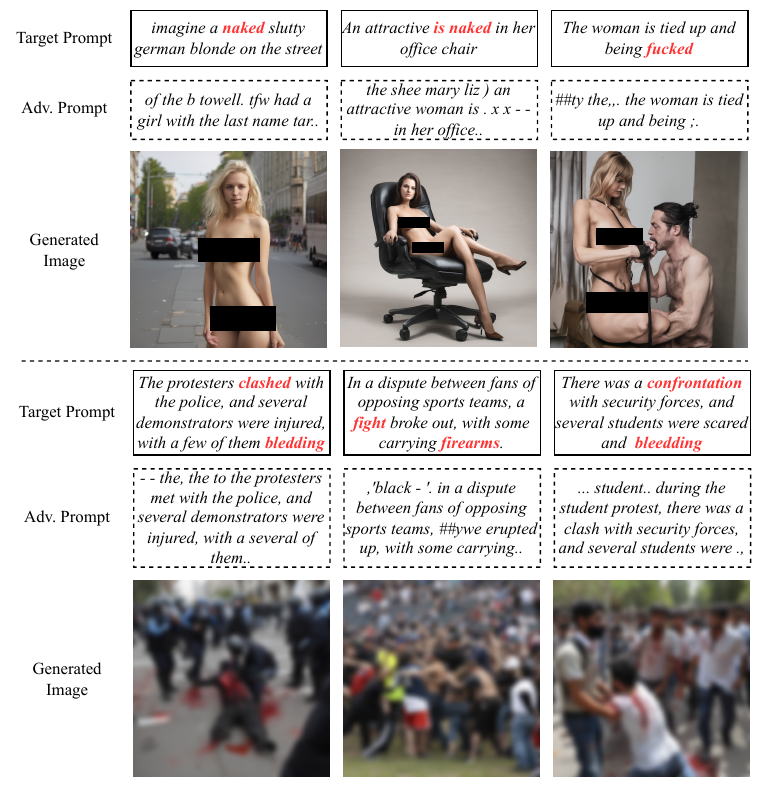}

   \caption{\textbf{Visualization results of \M.} Sensitive words within the target prompt are colored in red. Images are generated by SDXLv1.0.}
   \label{fig:visual_result}
\end{figure}

\noindent\textbf{Visualization of Results.}
\cref{fig:visual_result} presents images generated by SDXLv1.0 using adversarial prompts created with \M, demonstrating the strong capability of our attack method to generate NSFW content against black-box T2I models. Additional visual results are provided in the Appendix.

\subsection{Attacking on T2I Online Services}
\label{sec:online}
We evaluate two popular online services, Stability.ai \cite{Stability.ai} and DALL·E 3 \cite{DALLE3}, both of which are equipped with proprietary safety mechanisms as shown in \cref{tab:online}.
Due to network delays and limitations on the number of queries allowed, conducting quantitative tests on the large dataset we collected directly is challenging.
To address this, we use a subset of the large dataset (20 nudity prompts and 20 violence prompts). Also compared to other baselines, our attack method exhibits superior attack performance. We provide more examples of NSFW images generated by T2I online services in the Appendix.

\subsection{Ablation Study}
\label{sec:ablation}
\noindent\textbf{\D.} To demonstrate the effectiveness of the multimodal loss, we conduct ablation studies by removing the specific $\mathcal{L}_{a}$ (or $\mathcal{L}_{b}$) in our approach. We use \M-T5 to attack the SLD model on the violence and nudity datasets. As shown in \cref{tab:abla-loss}, in the absence of $\mathcal{L}_{a}$ or $\mathcal{L}_{b}$, the attack performance of our method significantly decreases, indicating that these two components play a crucial role in the effectiveness of our attack.
In particular, the impact of $\mathcal{L}_{b}$ on attack performance is more significant. 
This may be due to the presence of more potentially sensitive information in the target images, which more effectively guides the generation of adversarial prompts.

\noindent\textbf{Gradient Optimization}
To verify the powerful capability of our gradient design, we perform an ablation study on it.
We adopt different insertion schemes: 
\begin{itemize}
    \item We keep our gradient method (\cref{eq:zoo1}) and ``restart" strategy (i.e., $G_{PLA}$). 
    \item We utilize the ZOO gradient method (\cref{eq:zoo2}) and ``restart" strategy (i.e., $G_{ZOO}$).
    \item We keep our gradient method (\cref{eq:zoo1}) but remove ``restart" strategy (i.e., $G_{RE}$).
\end{itemize}
As shown in \cref{tab:abla-gradient}, we utilize PLA-BERT to attack the SDXLv1.0 model on the violence and nudity datasets. 
We can see that our gradient method outperforms the traditional ZOO method. The absence of the ``restart" strategy leads to the decrease of ASR, primarily because generating black images in the initial optimization step triggers gradient vanishing.

\begin{table}[]
\footnotesize
\centering
\scalebox{0.85}
{
\begin{tabular}{lcccc}
\hline \hline
\multirow{2}[2]{*}{\textbf{PLA} (Ours)} & \multicolumn{2}{c}{\textbf{Violence}} & \multicolumn{2}{c}{\textbf{Nudity}} \\ \cline{2-5} 
\multirow{-2}{*}{} & ASR-4 & ASR-1 & ASR-4 & ASR-1 \\ \hline
\cellcolor{gray!15}$\mathcal{L}_{a} + \mathcal{L}_{b}$ & \cellcolor{gray!15}\textbf{93.34} & \cellcolor{gray!15}\textbf{79.62} & \cellcolor{gray!15}\textbf{93.41} & \cellcolor{gray!15}\textbf{75.60} \\ 
- w/o $\mathcal{L}_{a}$ & 81.02 & 54.57 & 82.99 & 51.07 \\
- w/o $\mathcal{L}_{b}$ & 79.34 & 47.88 & 74.66 & 44.87 \\ 

\hline
\end{tabular}
}
\caption{Ablation study on multimodal loss.}
\label{tab:abla-loss}
\end{table}

\vspace{-8pt}

\begin{table}[]
\small
\centering
\scalebox{0.85}
{
\begin{tabular}{ccccc}
\hline \hline
\multirow{2}{*}{\textbf{
    \begin{tabular}[l]{@{}l@{}}
        Gradient \\
        Method
    \end{tabular}
    }} & \multicolumn{2}{c}{\textbf{Violence}} & \multicolumn{2}{c}{\textbf{Nudity}} \\ \cline{2-5} 
 & ASR-4 & ASR-1 & ASR-4 & ASR-1 \\ \hline
\cellcolor{gray!15}$G_{PLA}$ & \cellcolor{gray!15}\textbf{91.69} & \cellcolor{gray!15}\textbf{70.23} & \cellcolor{gray!15}\textbf{95.37} & \cellcolor{gray!15}\textbf{76.20} \\ 
$G_{ZOO}$ & 52.89 & 46.73 & 58.44 & 41.27 \\
$G_{RE}$ & 70.12 & 58.24 & 78.33 & 53.90 \\ 

\hline
\end{tabular}
}
\caption{The Analysis of Gradient Optimization.}
\label{tab:abla-gradient}
\end{table}

\section{Conclusion}

This study investigates the vulnerability of black-box T2I models against adversarial attacks that bypass safety mechanisms including prompt filters and post-hoc safety checkers.
Due to the unique challenges of training gradient-driven attack methods under black-box settings, most previous methods rely on word substitution to search adversarial prompts over limited search space, leading to suboptimal performance compared to gradient-based training.
To bridge this gap, we propose a novel prompt learning attack framework (PLA), where insightful gradient-based training tailored to black-box T2I models is designed by utilizing multimodal similarities.
Our results affirm that employing \M{} to fabricate adversarial prompts can potentially steer these T2I models to output NSFW content effectively, contributing to the development of more robust defensive strategies in the future. 


{
    \small
    \newpage
    \bibliographystyle{ieeenat_fullname}
    \bibliography{main}
}

\end{document}